\documentclass[manuscript]{aastex}
\usepackage{amsfonts}
\usepackage{amssymb}

\slugcomment{Not to appear in Nonlearned J., 45.}

\shorttitle{Collapsed Cores in Globular Clusters}
\shortauthors{Djorgovski et al.}

\begin{document}

\title{Electron Cyclotron Maser Emissions from Evolving Fast Electron Beams}

\author{J. F. Tang\altaffilmark{1,2}, D. J. Wu\altaffilmark{3}, L. Chen\altaffilmark{3}, G. Q. Zhao\altaffilmark{4} and C. M. Tan\altaffilmark{2}}

\email{jftang@xao.ac.cn}

\altaffiltext{1}{Xinjiang Astronomical Observatory, Chinese Academy of Sciences, 150 Science 1-Street, Urumqi, Xinjiang 830011, China}

\altaffiltext{2}{Key Laboratory of Solar Activity, National Astronomical Observatories, Chinese Academy of Sciences, Beijing 100012}

\altaffiltext{3}{Purple Mountain Observatory, CAS, Nanjing 210008, China}

\altaffiltext{4}{Institute of Space Physics, Luoyang Normal University, Luoyang 471022, China}

\begin{abstract}
Fast electron beams (FEBs) are common products of solar active phenomena. Solar radio bursts are an important diagnostic tool in the understanding of FEBs as well as the solar plasma environment in which they are propagating along solar magnetic fields. In particular, the evolutions of the energy spectrum and velocity distribution of FEBs due to the interaction with the ambient plasma and field when propagating can significantly influence the efficiency and property of their emissions. In this paper, we discuss some possible evolutions of the energy spectrum and velocity distribution of FEBs due to the energy loss processes and the pitch-angle effect caused by the magnetic field inhomogeneity, and analyze the effects of these evolutions on electron cyclotron maser (ECM) emission, which is one of the most important mechanisms of producing solar radio bursts by FEBs. The results show that the growth rates all decrease with the energy loss factor $Q$, but increase with the magnetic mirror ratio $\sigma$ as well as with the steepness index $\delta$. Moreover, the evolution of FEBs also can significantly influence the fastest growing mode and the fastest growing phase angle. This leads to the change of the polarization sense of ECM emission. In particular, our results also reveal that FEB that undergoes different evolution processes will generate different types of ECM emission. We believe the present results to be very helpful on more comprehensive understanding of dynamic spectra of solar radio bursts.

\end{abstract}


\section{INTRODUCTION}
Beams of energetic particles are the common products in various active phenomena in space and cosmic plasmas. For the Sun, particle acceleration is most prominent in flares and coronal shock wave. It is generally recognized that the magnetic reconnection process during flares can convert magnetic energy into thermal and kinetic energies of the accelerated particles \citep{masuda94, yokoyama01, hara11, imada13}. \citet{lin76} pointed out that as much as half of the liberated energy from flares can be converted into acceleration of charged particles through the magnetic reconnection process. Collisionless shocks are also the strong sources of fast electron beams (FEBs) \citep{blandford87}. For the coronal shock wave, there are two different generation mechanisms: blast waves are initiated by the plasma pressure of flare and piston-driven shock waves are due to coronal mass ejections (CMEs) \citep{nindos11}. The primary theory for explaining particles acceleration in the vicinity of shock waves is diffusive shock acceleration (DSA) \citep{bell78, drury83, blandford87}. Electron accelerated by the quasi-perpendicular shocks has been discussed by many authors and proposed that they can be accelerated through shock drift acceleration \citep{wu84, park12, guo15}. For the quasi-parallel shocks, when large amplitude magnetic fluctuation is present, the shocks can accelerate electrons efficiently \citep{masters13}.

FEBs reveal themselves not only in hard X-ray and $\gamma$-ray, but also in the radio emission \citep{pick90}. The most direct observational evidence in radio emission for FEBs is the type III bursts and decimetric blips, which have similar properties as type III events \citep{stahli87}. Type III bursts are due to FEBs traveling upward along the open magnetic field structures. Observations of radio bursts also reveal that FEBs also propagate in the downward direction (reverse-slope drift bursts) or along the closed magnetic field lines (type J and U bursts) \citep{aschwanden95}. When electron beams propagate downward along the closed loop, electrons with large pitch angles can be reflected and captured in the loop, while electrons with small pitch angles will precipitate into the denser chromosphere. It is the precipitate and trapped electrons which produce hard X-ray and radio emission, respectively. The key issue is how a beam of fast electrons leads to the generation of solar radio bursts described above. \citet{ginzburg58} proposed the so-called plasma emission. \textbf{In this conventional theory, Langmuir waves play a key role. FEBs accelerated by the magnetic reconnection process or coronal shock wave can excite Langmuir waves due to beam-plasma instability along their path when they propagate in corona. Then these electrostatic waves are partly converted into electromagnetic waves via nonlinear wave-wave interaction \citep{robinson93, wu94}. \citet{reiner98} found that plasma emission is the emission mechanism of type II bursts which is generated in the upstream region of the CME-driven shock. This theory which treats the acceleration of electrons near the shock, formation of FEBs, generation of Langmuir waves, and conversion of Langmuir waves into electromagnetic waves has been studied by many authors \citep{knock01, knock05, schmidt08}.} Another important coherent theory was suggested by \citet{twiss58} and \citet{schneider59} which direct amplify the electromagnetic waves at frequencies near the electron gyrofrequency and its harmonics. The electron-cyclotron maser (ECM) instability due to the wave-particle interaction is the amplification mechanism.

As a dominant mechanism for radio emission in astrophysics, ECM emission has been extensively applied to various short-time radio bursts from the magnetized planets, the Sun, and other stars \citep{treumann06}. In the references about ECM emission, the FEBs responsible for the generation of radio radiation are treated as an invariable source. However, FEBs propagating downwards deeper into the solar atmosphere will lose some of their energy via the interaction with the dense plasma \citep{pick90} and the energy spectrum of FEBs will evolve due to this energy loss process. High resolution X-ray observations suggest that the noncollisional energy loss of FEBs which are traveling in the loop did can flatten the spectrum of footpoint source \citep{battaglia08}. Furthermore, the velocity distribution of FEBs and the parameters of ambient plasma will change as they traveling in the complex magnetized plasma. In particular, the evolutions of the energy spectrum and velocity distribution of FEBs due to the energy loss processes and the pitch-angle scatter in the inhomogeneous magnetic field can significantly influence the efficiency and property of their emissions. Study of X-ray emission can provide us information about the acceleration and evolution processes of FEBs. On the other hand, X-ray emission has close association with solar radio emission \citep{kane81, aschwanden85, aschwanden92}, which indicates that X-ray and radio emissions are excited by the same evolving electron beam \citep{kundu82, gary85, aschwanden02}.

In this paper, we concentrate on the characteristic of ECM emission excited by the evolutional FEBs traveling in a coronal loop associated with CME. We suggest that the energetic electrons which excite the ECM emission are accelerated in the CME-driven shock front. The results show that the evolution of electron beams have significantly influence on ECM emission. With different energy loss process, the FEBs can produce different radio emissions. The paper is organized as follows. Firstly, we introduce the basic physical model in Section 2. Secondly, the ECM instability excited by the energetic electrons which travel in the loop is discussed in Section 3. Finally, discussions and conclusions are presented in Section 4.

\section{THE PHYSICAL MODEL}
\subsection{Magnetic Field Configuration of Source Region}
Beams of energetic electron are prominently created during solar flares by the magnetic reconnection process or during CMEs by the CME-driven shock wave. CMEs are defined as the outward traveling bright arc seen in the coronagraphs and the dark cavity \citep{forbes00}. It constitutes large-scale ejections of magnetic flux and mass and conceives a steady expansion as it propagating from lower corona to the interplanetary space, with two foot points on the Sun \citep{chen97}. High resolution X-ray imaging observation shows that all the CMEs and prominence eruptions have a giant arches structure \citep{svestka97, forbes00}. Here we consider the latter case and propose the magnetic field configuration of source region as Figure1. When the outward propagating CMEs overtake the local fast magnetosonic wave, it can drive a forward shock ahead of CME \citep{stewart74}. The CME-driven shock front is the acceleration site for solar energetic electrons. The acceleration mechanism of electrons at quasi-perpendicular shocks is the shock-drift acceleration, it is simple and effective \citep{wu94, park12, guo15}.

\begin{figure}
\epsscale{.70} \plotone{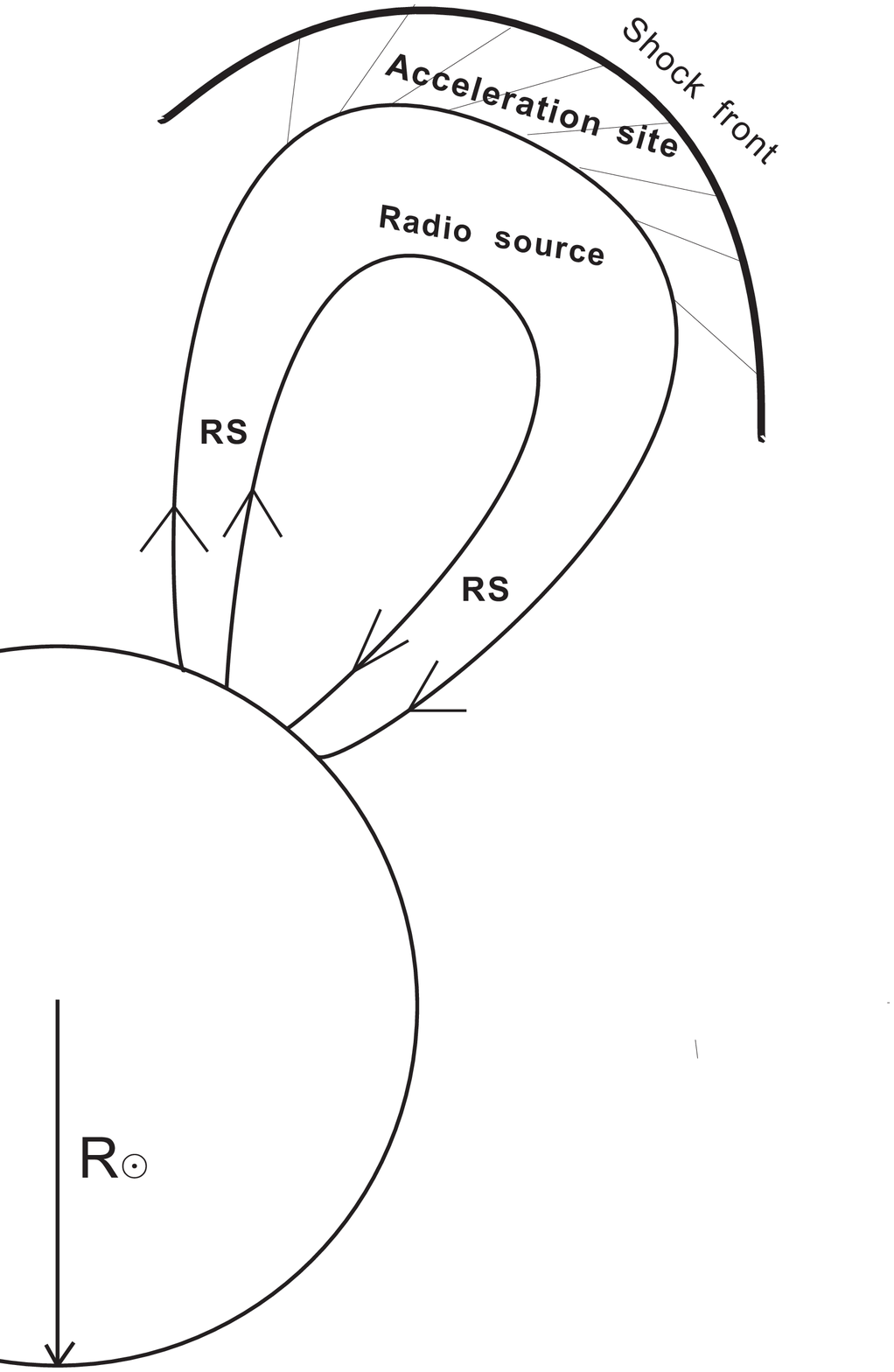} \caption{Schematic diagram of magnetic field configuration in the source region. The energetic electrons are accelerated in the CME-driven shock front. It is suggested that energetic electrons travel in the loop excite cyclotron maser emission and produce the radio emission.\label{fig1}}
\end{figure}

\citet{zhao14} proposed a model for the generation of radio radiation from the shock front. In their model, Alfv\'{e}n wave is first excited by the energetic ion beam accelerated in the shock front, then a density-depleted duct can form due to the excited Alfv\'{e}n wave along the foreshock boundary. Energetic electrons are also accelerated by the shock waves and part of the energetic electrons propagate in this density-depleted duct which drive the maser emission may give rise to type II bursts \citep{zhao14}. In our model, we also proposed that the FEBs which excite radio emission are accelerated by CME-driven shock wave.

When the energetic electrons leave the accelerate site, some of them propagate along the open magnetic field line, and the remaining energetic electrons will propagate along and be trapped in the magnetic loop. When the electron beam precipitates from the top to the footpoint they will lose most of their energy due to the interaction with the ambient plasma \citep{xu13}. The energy loss would result in a flatter spectral index and change the lower cutoff energy of the footpoint source. On the other hand, because of the complex magnetic field topology, velocity distribution of FEBs and the parameters of ambient magnetic plasma also will change when they propagating from the looptop to the footpoint. Therefore, the distribution functions of FEBs from the looptop to the footpoint are variational. It is these evolving FEBs with different evolution process excite cyclotron maser instability and produce different radio burst events.

\subsection{Evolution of Fast Electron Beams}
Energetic streams of electrons are created during various active phenomena. The most direct observational evidence for energetic electrons are from X-ray, $\gamma$-ray, and radio emission. When the energetic electrons escape from the acceleration site, some of them move outwards into the interplanetary space and often can be detected via their characteristic emission of corona and interplanetary type III bursts. The others propagate downwards along the magnetic field lines, because of the convergent magnetic field, electrons with small pitch angles will precipitate into the dense chromosphere and electrons with large pitch angles will be reflected at the mirror points. It is this precipitate and reflected electrons produce the X-ray and radio emissions. Observations frequently demonstrate that there is a looptop X-ray source in the corona and two or more footpoint X-ray sources in the chromosphere \citep{frost71, hudson78, hoyng81}. \citet{kosugi88} made a statistical study and show that the X-ray has a strong association with the microwave emission. This indicates that they could be excited by the same energetic electron beam in the different levels of solar atmosphere.

Electron beams traveling in a closed loop will loose most of their energy. High resolution observations of X-ray emissions and the evolution of X-ray spectrum could provide us information for understanding the physics of acceleration and energy loss of the FEBs \citep{zharkova95, hannah11, xu13}. It is believed that both the looptop and the footpoint X-ray emissions are produced by a single FEB which travels along the loops structure from the looptop to footpoint. The emission mechanism for the X-ray source is the thin-target bremsstrahlung at the looptop and the thick-target bremsstrahlung at the footpoints, respectively. According to this emission model, \textbf{the accelerated electron beam with power-law spectrum distribution $F(E)\propto E^{-\delta_b}$ in the corona will emit X-ray emission at the looptop source with the photon spectrum $I(\epsilon)\propto \epsilon^{-\gamma_{lt}}$, here the spectral index $\gamma_{lt}=\delta_b+1$, $E$ is the energy of electron, $\epsilon$ is the energy of photon} \citep{arnoldy68, lin76, hannah11}. If the power-law electrons only encounter Coulomb collisions with the ambient plasma when they propagate to the chromosphere, the footpoint source could produce a power-law spectrum $I(\epsilon)\propto \epsilon^{-\gamma_{fp}}$ with an index of $\gamma_{fp}=\delta_b-1$ \citep{brown71, hannah11}. This standard flare model indicates the difference of the spectral indices between the looptop and the footpoint sources is 2 \citep{hannah11}. However, observations of solar flare demonstrate that the difference of spectral index is inconsistent with the prediction of emission models \citep{masuda94, masuda95}. \citet{petrosian02} proposed that the average difference is about $\Delta\gamma=1.3\pm1.5$, according to the survey of Yohkoh flares observations. This discrepancy between observations and predictions suggests that there must exist other energy loss mechanism except the Coulomb collisions. \citet{battaglia08} proposed that noncollisional energy loss via the induced electric field can flat the footpoint spectrum and lead to the larger difference of spectral indices. This induced electric field is due to the return current generated by the energetic electron beam. \citet{hannah11} investigate the spectral evolution of X-ray sources including wave-particle interactions and found that the growth of Langmuir waves can flatten the spectrum of footpoint source. The looptop source is unchanged and so the difference of spectral indices can be greater than 2. To study the evolution of energetic electron spectrum travelling in the solar atmosphere, pitch angle scattering and various energy loss mechanisms, such as Coulomb collisions, anomalous resistivity in the return current system, wave-particle interactions and so on should be taken into account.

For simplicity, we assume that the energy loss of energetic electrons, $\delta E$ is independent of their initial energy $E$ when electrons travel in the loop structure. So the energy distribution function of energetic electrons for the initial form
\begin{equation}
F_{lt}(E)=AE^{-\delta}
\end{equation}
at the looptop will evolve into the final form because of the energy loss $\delta E$ \citep{xu13}:
\begin{equation}
F_{fp}(E')=A(E'+\delta E)^{-\delta}.
\end{equation}
Here A is the normalized factor, $E'=E-\delta E$ is the energy of energetic electrons when they arrive at the footpoint \citep{xu13}.

\citet{wang04} proposed that energetic electrons captured at the apex of the magnetic loop have a beam-like velocity distribution. In our model, electrons are accelerated in the advancing shock front. Some of the energetic electrons leave the acceleration region and precipitate along the arch structure (see Fig.1). So we also propose that the precipitated electrons have a beam velocity distribution when they leave the acceleration site. On the other hand, hard X-ray observations demonstrate that energetic electrons, which can excite ECM instability, generally have an approximate power-law energy distribution with a lower energy cutoff \citep{lin74, stupp00, aschwanden02}. The spectral index of power-law electrons typically $\alpha=3$ and is in the range of $2\sim6$ \citep{stupp00}. In principle, it is difficult to describe the special form of the lower energy cutoff behavior based on observations. \citet{wu08} fitted the more general power-law spectrum with the lower energy cutoff behavior described by a hyperbolic tangent function. Consequently, we propose the following electron distribution function when they leave the acceleration site:
\begin{equation}
F_0(u, \mu)=A_0tanh\left(\frac{u}{u_c}\right)^{2\delta}\left(\frac{u}{u_c}\right)^{-2\alpha}\exp\left[-\frac{(u\mu-u_c)^2}{\beta^2}-\frac{u^2(1-\mu^2)}{\beta^2}\right],
\end{equation}
here, $u^2=u_{\perp}^2+u_{\parallel}^2$, $\textbf{u}=\textbf{p}/m$ denotes the momentum per unit mass, $u_{\perp}$ and $u_{\parallel}$ are the perpendicular and parallel components of $\textbf{u}$ to the ambient magnetic field, respectively. Parameter $\mu=u_{\parallel}/u$, $A_0$ is the normalization coefficient. $\delta$ is the steepness index, $E_c=\frac{1}{2}mu_c^2$ describes the cutoff energy, and the hyperbolic tangent function $tanh(u/u_c)^{2\delta}$ describes the lower energy cutoff behavior. $\alpha$ is the spectrum index of the energetic electrons, and $\beta$ is the momentum dispersions in $u_{\perp}$ and $u_{\parallel}$.

Based on the assumption that the energy loss of FEBs $\delta E$ is independent of their initial energy, we can obtain the distribution function of FEBs after they leave the acceleration site as below:
\begin{eqnarray}
F_1(u_1, \mu) &= &
A_0tanh\left(\frac{u_1^2}{u_c^2}+Q\right)^{\delta}\left(\frac{u_1^2}{u_c^2}+Q\right)^{-\alpha} \nonumber\\
 & \times &\exp\left[-\frac{\left(\mu\sqrt{\frac{u_1^2}{u_c^2}+Q}-1\right)^2}{\beta'^2}-\frac{\left(\frac{u_1^2}{u_c^2}+Q\right)(1-\mu^2)}{\beta'^2}\right].
\end{eqnarray}
Where, $Q=\delta E/E_c$, $\beta'=\beta/u_c$. Then, taking into account the magnetic mirror due to the magnetic field convergence at the footpoints, we have the distribution function of FEBs as:
\begin{eqnarray}
F_1(u_1, \mu) &= &
A_0tanh\left(\frac{u_1^2}{u_c^2}+Q\right)^{\delta}\left(\frac{u_1^2}{u_c^2}+Q\right)^{-\alpha}\left[1-\exp\left[(1-\sigma)\frac{1-\mu^2}{\mu^2}\right]\right] \nonumber\\
& \times &\exp\left[-\frac{\left(\mu\sqrt{\frac{u_1^2}{u_c^2}+Q}-1\right)^2}{\beta'^2}-\frac{\left(\frac{u_1^2}{u_c^2}+Q\right)(1-\mu^2)}{\beta'^2}\right].
\end{eqnarray}
Here, $\sigma$ is magnetic mirror ratio parameter.

\section{NUMERICAL RESULTS}
\subsection{ECM Emission Theory}
ECM emission is a well-known radiation mechanism and has been extensively applied to various solar radio bursts \citep{treumann06}. Based on the cold-plasma theory, we have the dispersion relation of high frequency electromagnet emission as \citep{wu02}:
\begin{equation}
N_q^{2}=1-{\omega_p^2\over \omega\left(\omega+\tau_q\omega_{ce}\right)},
\end{equation}
here, $\omega_p$ is the plasma frequency of ambient plasma, $\omega_{ce}$ is the electron cyclotron frequency, $N_q$ and $\omega$ are the refractive index and frequency of the excited wave, respectively. $\tau_q=-s_q+q\sqrt{s_q^2+\cos^2\theta}$, $s_q=\omega\omega_{ce}\sin^2\theta/2\left(\omega^2-\omega_p^2\right)$, and $\theta$ is the wave phase angle with respect to ambient magnetic field, $q=\pm$ denote the ordinary (O) and extraordinary (X) modes, respectively. When frequency of excited wave $\omega\simeq s\omega_{ce}$, the temporal growth rate of the excited wave can be given by follow form \citep{wu02}:
\begin{eqnarray*}
\omega_{qi} & = & {\pi\over 2}{n_b\over n_0}{\omega_{p}^2\over\omega}\int
d^3{\bf
u}{\gamma\left(1-\mu^2\right)\over\left(1+T_q^2\right)R_q}
\delta(\gamma-{s\omega_{ce}\over\omega}-{N_q\mu u\over c}\cos\theta) \nonumber \\
 & &\times\left\{{\omega\over\omega_{ce}}\left[\gamma K_q\sin\theta+T_q\left(\gamma\cos\theta-{N_q\mu u\over c}\right)\right]{J_s(b_q)\over b_q}
 \right. \nonumber
\end{eqnarray*}
\begin{equation}
\qquad{}
 \displaystyle\biggl.+J_s^\prime(b_q)\biggr\}^2\left[u{\partial\over \partial u}+\left({N_qu\cos\theta\over c\gamma}-\mu\right){\partial\over \partial\mu}\right]F_b(u,\mu).
\end{equation}
Here,
\begin{eqnarray}
b_q & = & N_q{\omega\over\omega_{ce}}{u\over c}\sqrt{1-\mu^{2}}\sin\theta, \nonumber \\
R_q & = & 1-{\omega_{p}^2\omega_{ce}\tau_q\over 2\omega\left(\omega+\tau_q\omega_{ce}\right)^2}
\left(1-{q s_q\over\sqrt{s_q^2+\cos^2\theta}}{\omega^2+\omega_{p}^2\over\omega^2-\omega_{p}^2}\right), \nonumber \\
K_q & = &
{\omega_{p}^2\omega_{ce}\sin\theta\over\left(\omega^2-\omega_{p}^2\right)\left(\omega+\tau_q\omega_{ce}\right)},
\quad T_q=-{\cos\theta\over\tau_q}.
\end{eqnarray}

Above, $n_b$ and $n_0$ are the number densities of the energetic electron beam and ambient plasma. $J_s(b_q)$ is the Bessel function, and $\gamma=\sqrt{1+(u/c)^2}$ is the Lorentz factor.

\subsection{The Growth Rate of ECM Instability}
With the velocity distribution function $F_1(\mu,u_1)$ given by equation(5), the growth rates of ECM emission in the O and X modes can be calculated based on equation(7). For given parameters ($\delta, \alpha, u_c, \beta$, and Q) of FEBs and background magneto-plasma parameters ($\Omega$ and $\sigma$) in the source region, the growth rates of ECM instability depends on parameters $\omega$ and $\theta$. Here $\Omega$ is the frequency ratio of $\omega_{ce}$ to $\omega_p$. Both the peak and the maximum growth rates are normalized by $\omega_{ce}n_b/n_0$. Figure 2 presents the peak growth rates with varying $\omega$ and fixed phase angle $\theta$. Figure 3 shows the maximum growth rates with varying $\omega$ and $\theta$ together, versus frequency ratio $\Omega$. Panels O1 and X2 denote the fundamental wave of O mode and harmonic wave of X mode, respectively. The solid line, dotted line, and dot dash line denote the growth rates of ECM instability excited at looptop (LT), loopmid (LM), and footpoint (FP) sources, respectively. Here the spectrum index $\alpha=3$, deepness index $\delta=6$, $u_c=0.3c$, $\beta=0.25c$, and for Figure 2, frequency ratio $\Omega=2$ has been used. When the FEBs travel in the loop from the top to the footpoint, parameters $\sigma=0, Q=0.1$; $\sigma=0, Q=0.51 (O1)$ and $Q=0.7 (X2)$; and $\sigma=3.5, Q=1.5$ have been used for the LT, LM, and FP sources, respectively. Both in Figure 2 and 3, the growth rates of LM source have been enlarged by a factor of 100. This implies that the growth rates at LT and FP sources are at least two orders of magnitude greater than that of LM source.
\begin{figure}
\epsscale{.7} \plotone{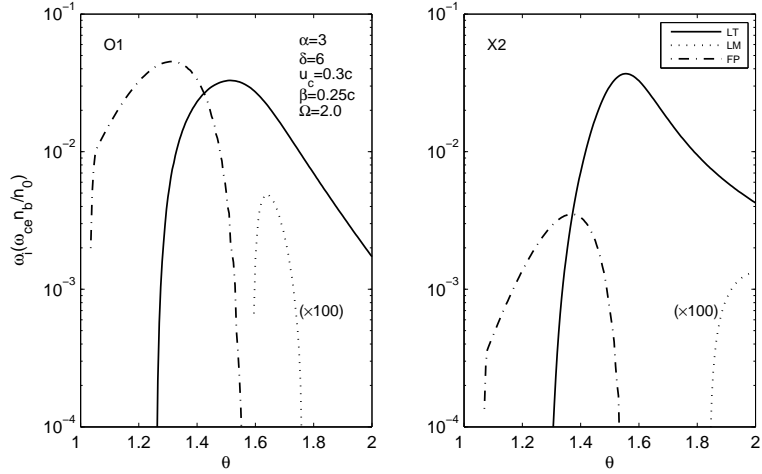}
\caption{\textbf{The peak growth rates of O1 and X2 modes excited by the FEBs which travel in the loop. The deepness index $\delta=6$. It shows that the growth rates of LM source are two orders of magnitude smaller than that of the LT and FP sources.} \label{fig2}}
\end{figure}

\begin{figure}
\epsscale{.7} \plotone{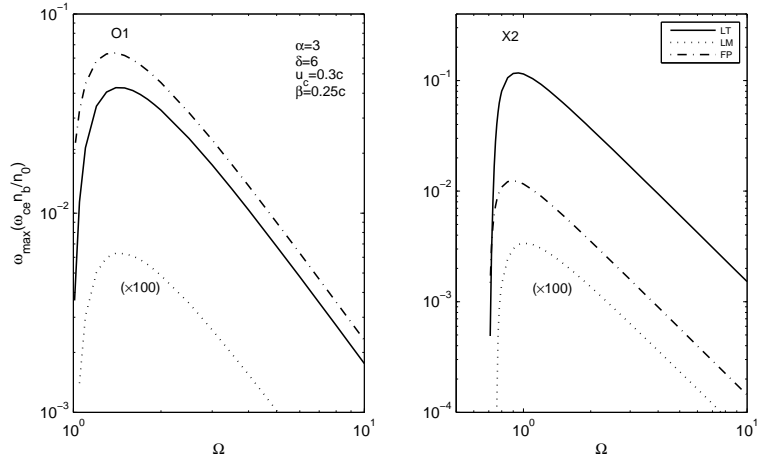}
\caption{\textbf{The maximum growth rates of O1 and X2 modes as a function of the plasma parameter $\Omega$. The deepness index $\delta=6$. The growth rates of LM source have been enlarged by a factor of 100.} \label{fig3}}
\end{figure}

It is found that the energy loss process of FEBs can significantly influence the efficiency and property of ECM emission. The growth rates of O1 and X2 modes all decrease rapidly with the energy loss factor $Q$. Figure 2 shows that the phase angles $\theta$ of the maximum growth rates of LM source can deviate from that of LT and FP sources obviously due to the energy loss of fast electrons. The velocity distribution of FEBs will evolve due to the magnetic field inhomogeneity and also influence the efficiency and property of their emission. Our calculation show that the growth rates of FP sources can increase to a comparable value to the LT sources because of the convergent magnetic field at footpoint. Figure 2 and 3 show that the O1 mode has larger growth rates than X2 mode and becomes the dominant mode at FP sources due to the loss-cone anisotropy ($\sigma=3.5$). Figure 2 and 3 imply that if the FEBs have a moderate steepness cutoff behavior ($\delta=6, \alpha=3$) when they leave the acceleration region and the loop have magnetic mirror effect only in the footpoint region, the growth rates at LM sources are very small. One can anticipate that the growth rates of LM sources will three or four orders of magnitude smaller than that of LT and FP sources if the FEBs loss more energy (i.e., $Q>0.51$ or 0.7) when they travel from top to middle part of the loop. So for the case of Figure 2 and 3, ECM emission from the evolving FEBs which travel in the loop will form three separate radio sources.

Figure 4 and 5 present the peak and maximum growth rates as a function of phase angle $\theta$ and frequency ratio $\Omega$, respectively. The acronym LT, LM, and FP also denote the looptop source, loopmid source, and footpoint source, respectively. The spectrum index $\alpha=3$, deepness index $\delta=10$, $u_c=0.3c$, $\beta=0.25c$, and $\Omega=2$ in Figure 4 have been used. Here, the magnetic mirror ratio $\sigma$ and energy loss factor Q have the same values as Figure 2 and 3 for the LT, LM and FP sources. It also shows that the O1 mode will has larger growth rates than X2 mode and becomes the dominant mode at FP sources. Compare Figure 2 with 4 one can find that the growth rates all increase with the steepness index $\delta$, especially for the LM sources. Figure 4 and 5 imply that if the energetic electrons have a steeper cutoff behavior (i.e., have a larger steepness index) when they leave the acceleration region, the growth rates of LM source are comparable to that of LT or FP sources, even larger than the growth rates of FP source (see Figure 5). So for this case, ECM emission excited by the evolving FEBs which propagate from the looptop to the footpoint will form a continuous radio source.

\begin{figure}
\epsscale{.70} \plotone{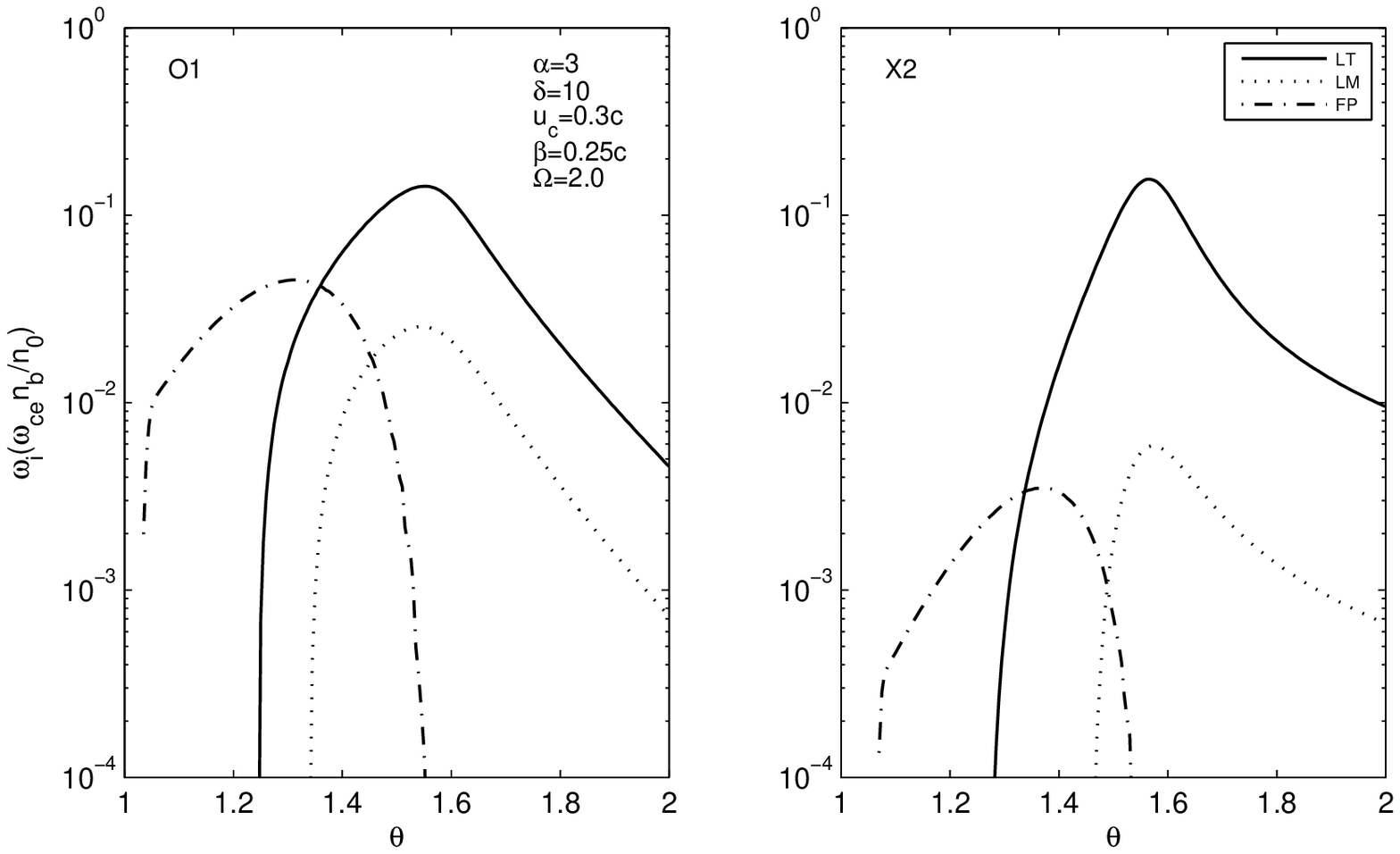}
\caption{\textbf{The peak growth rates of O1 and X2 modes excited by the FEBs which travel in the loop. The deepness index $\delta=10$. It shows that the growth rates of LM source are comparable to that of LT or FP sources.} \label{fig4}}
\end{figure}
\begin{figure}
\epsscale{.70} \plotone{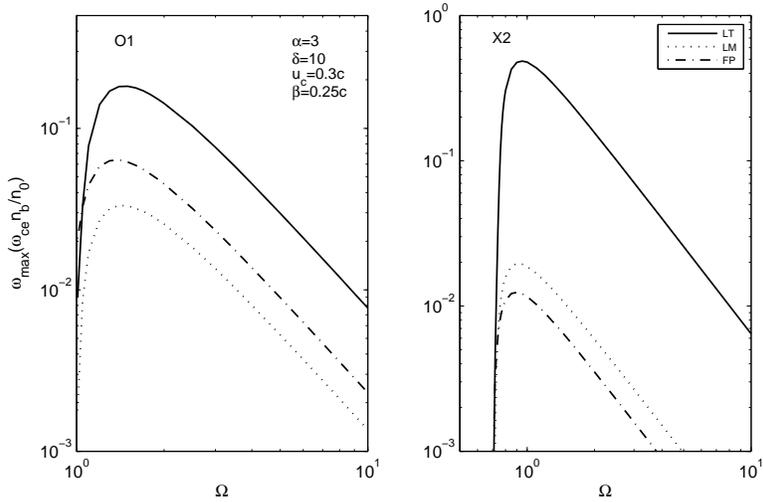}
\caption{\textbf{The maximum growth rates of O1 and X2 modes as a function of the plasma parameter $\Omega$. The deepness index $\delta=10$.} \label{fig5}}
\end{figure}

In Figure 6, we also plotted dependence of maximum growth rates on the frequency ratio $\Omega$, where the parameters $\alpha=3$, $\delta=6$, $u_c=0.3c$, and $\beta=0.25c$ have been used. The solid line, dotted line, and dot dash line also correspond to emission at LT, LM, and FP sources, respectively. Here, we choose the parameters $\sigma=0, Q=0.1$; $\sigma=1.5, Q=0.51 (O1), Q=0.7 (X2)$; and $\sigma=3.5, Q=1.5$ for the LT, LM, and FP sources, respectively. It shows that the growth rates of LM sources can be greater than that of LT or FP source if the magnetic field converge from the middle part of the loop and the dominant mode also shift from X2 mode to O1 mode at the LM and FP sources. Figure 6 implies that if the CME loop has a effective magnetic mirror effect from middle part, ECM emission from the evolving FEBs which travel and trapped in such magnetic loop will form a continuous radio burst.

\begin{figure}
\epsscale{.70} \plotone{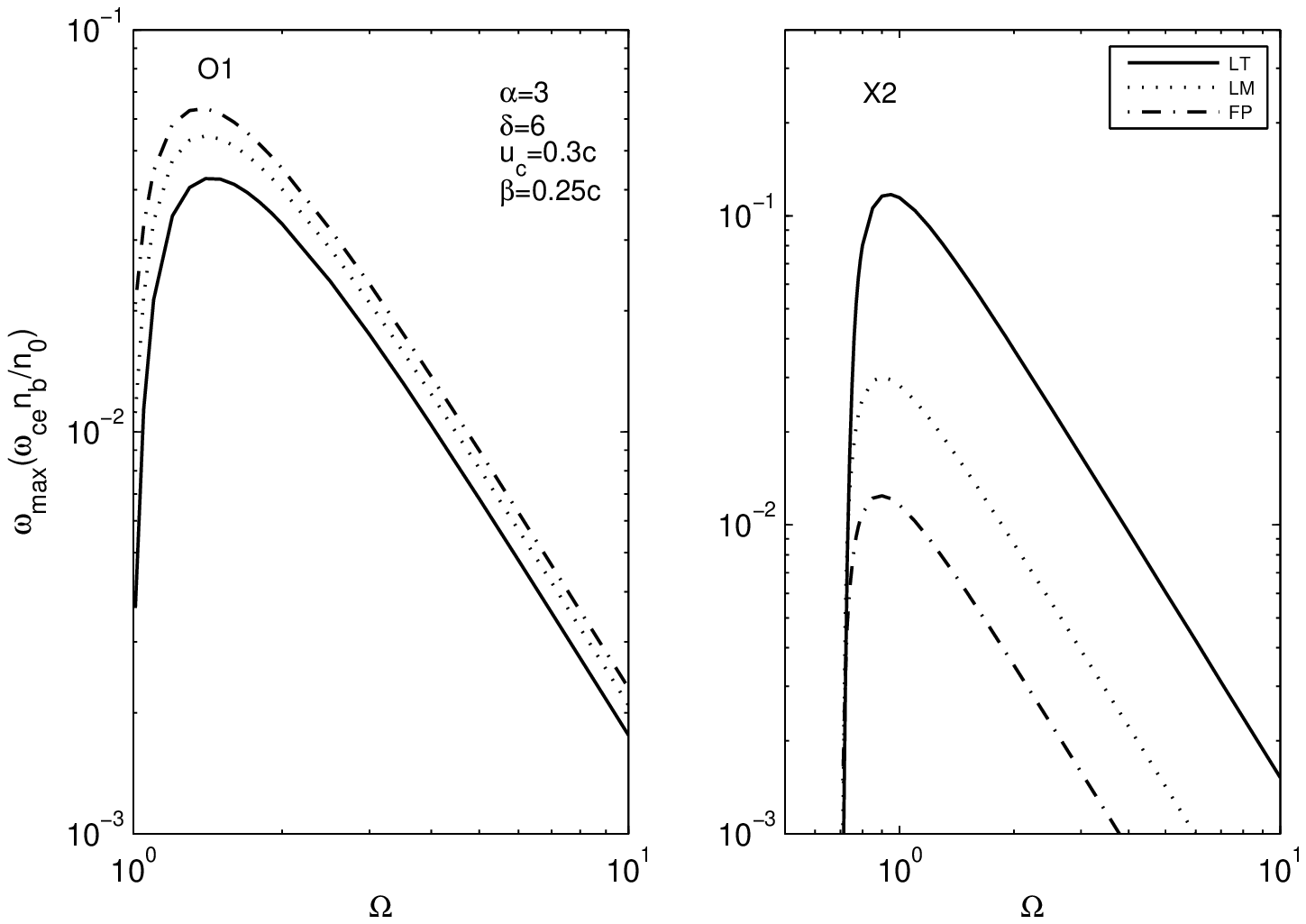}
\caption{\textbf{The maximum growth rates of O1 and X2 modes as a function of the plasma parameter $\Omega$. The deepness index $\delta=6$ and the magnetic field line converge from the middle part of the loop. The results also show that the growth rates of LM source are comparable to that of LT or FP sources.} \label{fig6}}
\end{figure}

\section{DISCUSSIONS AND CONCLUSIONS}
Particle acceleration is one of the most significant process which is ubiquitous in space and cosmic plasmas. For the Sun, it is generally acknowledged that FEBs usually are produced by the magnetic reconnection process during flares or by the diffusive shock acceleration in the vicinity of corona shock wave. FEBs can be detected directly by X-ray, $\gamma-$ray, and radio observations and we can diagnose the FEBs as well as the solar plasma environment from solar radio bursts. When the FEBs traveling in the loop will loss some of their energy due to various energy loss mechanism and get a more flat spectrum. The velocity distribution of FEBs and the magnetic plasma parameters will change as they traveling in the solar plasma. So it is significative to study the radio emission from evolving FEBs which are traveling in solar atmosphere.

In this paper, we discuss the possible evolution of the energy spectrum and the velocity distribution of FEBs and analyze the effects of these evolution on ECM emission. Results from our calculations show that the evolution of the energy spectrum and velocity distribution of FEBs can significantly influence the efficiency and property of ECM emissions. It is clear that the growth rates all decrease rapidly with the energy loss factor $Q$ but increase with the steepness index $\delta$ and magnetic mirror ratio $\sigma$, respectively. The growth rates of O1 mode can be larger than that of X2 mode and becomes the dominant mode at FP and LM sources due to the energy loss and the evolution of velocity distribution of FEBs. The results also show that the phase angles $\theta$ of the maximum growth rates of LM source can deviate from that of LT and FP sources obviously when the fast electrons lose most of their energy and the magnetic field don't converge at the LM source. If the FEBs have a moderate steepness cutoff behavior ($\delta=6, \alpha=3$) and the loop have magnetic mirror effect only in the footpoint region, the growth rates at LM sources are very small and the emission from such evolving FEBs will form three separate radio sources. If the FEBs have a steeper cutoff behavior (i.e., have a larger steepness index) or the loop has a effective magnetic mirror effect from middle region, the growth rates of LM sources are comparable to that of LT and FP sources and the emission excited by such evolving FEBs will form a continuous radio source.

The moving type IV (IVM) burst is a relatively rare form and it occurs after a solar flare or a coronal mass ejection. Sine the identification of IVM bursts, plenty of work have been done to explain the emission process of this rare bursts. It is widely believed that IVM bursts have a strong association with radio type II bursts \citep{boischot57, smerd71}. \citet{kai69} proposed a common shock wave as the source for generation of IVM and type II bursts. However, the radiation mechanism of the two types of bursts remains as a open problem. Based on our proposed model, cyclotron maser mechanism can be the possible emission process of IVM bursts. Energetic electron beams accelerated in the CME-driven shock front and excite ECM emission along the expanding CME associated loop. If the accelerated electrons have a moderate steepness cutoff behavior and the expanding loop have magnetic mirror effect only in the footpoint, ECM emission can form three separate radio sources, i.e., the expand arch IVM bursts. If the energetic electron beams have a steeper cutoff behavior when they leave the acceleration site or the CME loop has a effective magnetic mirror effect from the middle part, these trapped electrons will excite ECM emission and form a continuous radio source, i.e., the advancing front IVM bursts.

\acknowledgments This work is supported by NSFC under grant 11303082, 41531071, 11373070, 41304136, 41504131 and 11403087, by the West Light Foundation of CAS (No.XBBS201223) and by Key Laboratory of Solar Activity at National Astronomical Observatories, CAS.

\end{document}